\documentclass{siamltex}

\usepackage{epsfig}

\newtheorem{definitions}[theorem]{Definitions}
\newtheorem{observation}[theorem]{Observation}

\newcommand{\lowm}{{m_{\min}}}
\newcommand{\highm}{{m_{\max}}}
\newcommand{\depth}{\mbox{depth}}
\newcommand{\parent}{\mbox{parent}}
\newcommand{\child}[2]{\mbox{child}_{#1}(#2)}

\newcommand{\length}[1]{c_{#1}}

\newcommand{\Sprout}{\mbox{\sc Sprout}}
\newcommand{\Level}{\mbox{\sc Level}}
\newcommand{\UpdateHeaps}{\mbox{\sc Update-Qs}}
\newcommand{\AddTerminal}{\mbox{\sc Add-Terminal}}
\newcommand{\CreateTlowm}{\mbox{\sc Create-$T_{\lowm}$}}
\newcommand{\ComputeTrees}{\mbox{\sc Compute-Trees}}

\newcommand{\n}{\mbox{\bf N}}   
\newcommand{\C}{\mbox{\bf C}}   
\newcommand{\CMin}{\mbox{\bf C}_{\min}} 
\newcommand{\Deg}{\mbox{\bf mDeg}}
\newcommand{\M}{\mbox{\bf m}}   
\newcommand{\D}[1]{{\mbox{\bf D}[{#1}]}} 
\newcommand{\lowkey}[1]{{\mbox{\bf u}[{#1}]}}
\newcommand{\highkey}[1]{{\mbox{\bf w}[{#1}]}}
\newcommand{\lowheap}{\mbox{{\bf low}-queue}}
\newcommand{\highheap}{\mbox{{\bf high}-queue}}

\newenvironment{tabAlgorithm}[1]{
\setcounter{algorithmLineNum}{1}
\samepage
\begin{tabbing}
9\=999\=\kill #1
}{
\end{tabbing}
}
\newcounter{algorithmLineNum}
\newcommand{\algline}{\\\>\thealgorithmLineNum.\hfil\>
                        \stepcounter{algorithmLineNum}}
\newcommand{\algnono}{\\\> \>}
\newcommand{\CMT}[1]{---\= {\em #1} ---}

\title{Prefix Codes: Equiprobable Words, Unequal Letter Costs}

\author{Mordecai J. Golin\thanks{
    Hong Kong UST, Clear Water Bay, Kowloon, Hong Kong.
    Partially supported by HK RGC Competitive Research Grant HKUST 181/93E.
    Email: golin@cs.ust.hk}
  \and Neal Young\thanks{
    Dartmouth College, Hanover, NH, USA.
    Part of this work was done while at
    UMIACS, University of Maryland, College Park, MD 20742.
    Partially supported by NSF grants CCR--8906949 and CCR--9111348.
    Email: young@umiacs.umd.edu.}}

\begin{document}
\maketitle

\begin{abstract}
  We consider the following variant of Huffman coding in which the costs of the
  letters, rather than the probabilities of the words, are non-uniform: ``Given
  an alphabet of $r$ letters {\em of non-uniform length}, find a
  minimum-average-length prefix-free set of $n$ codewords over the alphabet;''
  equivalently, ``Find an optimal $r$-ary search tree with $n$ leaves, where
  each leaf is accessed with equal probability but the cost to descend from a
  parent to its $i$th child depends on $i$.''  We show new structural
  properties of such codes, leading to an $O(n\log^2 r)$-time algorithm for
  finding them.  This new algorithm is simpler and faster than the best
  previously known $O(nr\, \min\{\log n, r\})$-time algorithm due to Perl,
  Garey, and Even \protect{\cite{PGE-75}}.
\end{abstract}


\begin{AM}
  Analysis of Algorithms.
\end{AM}

\pagestyle{myheadings}
\thispagestyle{plain}
\markboth{M. GOLIN AND N. YOUNG}{PREFIX CODES}

\section{Introduction}

The well-known Huffman coding problem \cite{Huff-52} is the following:
given a sequence of access probabilities $\langle p_1, p_2, ..., p_n \rangle$,
construct a binary prefix code $\langle w_1, w_2, ..., w_n \rangle$
minimizing the expected length $\sum_i p_i \cdot \mbox{length}(w_i)$.
A {\em binary prefix code} is a set of binary strings,
none of which is a prefix of another.

A natural generalization of the problem is to allow the words of the code to be
strings over an arbitrary alphabet of $r \ge 2$ letters and to allow each
letter to have an arbitrary non-negative length.  The length of a codeword is
then the sum of the lengths of its letters.  For instance, the ``dots and
dashes'' of Morse code are a variable-length alphabet with length corresponding
to transmission time. (See Figure \ref{comparisons}.)  This generalization of
Huffman coding to a variable-length alphabet has been considered by many
authors, including Altenkamp and Mehlhorn \cite{AM-80}, and Karp
\cite{Karp-61}.  Apparently no polynomial-time algorithm for it is known, nor
is it known to be NP-hard.

A prefix code in which the codewords $\langle w_1, w_2, .., w_n \rangle$ are in
alphabetical order is called {\em alphabetic} \cite{AM-80}.  In this case the
underlying tree represents an $r$-ary {\em search tree}.  The length of the
$i$th letter corresponds to the time required to descend from a node into its
$i$th subtree.  This time is often a function of $i$ in search-tree algorithms,
for instance, when the subtree to descend into is chosen by sequential search.
An optimal alphabetic code thus corresponds to a minimum expected-cost search
tree.

In this paper we consider the special case in which the codewords occur with
equal probability, i.e., each $p_i$ equals $1/n$.  With this restriction, the
alphabetic and non-alphabetic problems are equivalent.  The problem may be
viewed as a variant of Huffman coding in which the lengths of the letters,
rather than the codeword probabilities, are non-uniform.  Alternatively, it may
viewed as the problem of finding an optimal $r$-ary search tree, where the
search queries are uniformly distributed but the time to descend from a parent
to its $i$th child depends on $i$.  For the complexity results stated in this
paper, the algorithms return a tree representing an optimal code.

In 1989, Kapoor and Reingold \cite{KR-89} 
described a simple $O(n)$-time algorithm for the binary case $r=2$.
In 1975, Perl, Garey, and Even \cite{PGE-75} 
gave an $O(rn\min\{r,\log n\})$-time algorithm.
(Although due a typographical error their abstract
incorrectly claims an $O(rn)$-time algorithm.)
In the same year Cot \cite{Cot-75} described an $O(r^2 n)$-time algorithm.
In 1971, Varn \cite{Varn-71} gave an algorithm
without analyzing its complexity.
It appears Varn's algorithm requires $\Omega(rn)$ time.

In this paper we describe an
$O(n\log^2 r)$-time algorithm based on new insights into the
structure of optimal trees.    
In Section 2 we define {\em shallow} and {\em proper} trees and
prove that some proper shallow tree is optimal.
In Section 3 we develop the algorithm, which efficiently constructs
all proper shallow trees and returns one representing an optimal prefix code.

\section{Shallow Trees}
\begin{figure}[t]
  \centerline{\psfig{figure=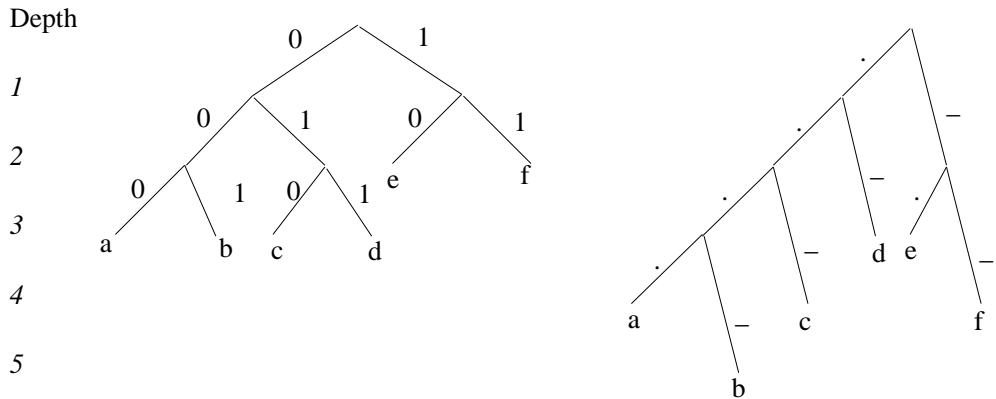,width=\hsize}}
  \caption{Two trees for the 6 symbols a,b,c,d,e,f,  each
    occuring with probability $1/6.$  The tree on the left is
    the optimal tree that uses the alphabet $\{0,1\},$ 
    $length(0)=length(1)=1$, while the tree on the right is for the alphabet
    $\{.,\_\}$ with $length(.)=1$ and $length(\_)=2.$
    The corresponding sets 
    of codewords are 
    $$a= 000,\quad b=001,\quad c=011,\quad d=011,\quad e=10,\quad f=11$$
    and 
    $$a = ....\,, \quad 
    b=...\_\,,    \quad 
    c=..\_\,,     \quad 
    d=.\_\,,      \quad 
    e=\_.\,,      \quad 
    f=\_\,\_$$} 
  \label{comparisons}
  \smallskip
  \hrule
  \smallskip
\end{figure}

Fix an instance of the problem,
given by the respective lengths 
$\langle \length{1} \le \length{2} \le \cdots \le \length{r} \rangle$
of the $r$ letters in the alphabet
and the number $n$ of (equiprobable and prefix-free) codewords required.
We assume the standard tree representation of prefix codes,
as described in the following definition.
\begin{definition}
  The {\em infinite $r$-ary tree} is the infinite, rooted, $r$-ary tree.
  Each tree edge has a length and a label ---
  an edge going from a node to its $i$th child has length $\length{i}$
  and is labeled with the $i$th letter in the alphabet.

  A {\em node} is a node of the infinite $r$-ary tree.
\end{definition}
The finite words over the alphabet of $r$ letters
correspond to the nodes.
The labels along the path from the root to any node 
spell the corresponding word
and the length of the path is the length of this word.
A prefix code corresponds to a set of nodes
none of which is a descendant of another. 
(See  Figure \ref{comparisons}.)

\begin{definitions}
  A {\em tree} is any subtree $T$ 
  of the infinite $r$-ary tree containing the root.
  In any tree, $n$ of the leaves will be identified as {\em terminals};
  their corresponding words form a prefix code.  
  The remaining nodes in the tree are referred to as {\em non-terminals}.

  Give a node $u$,
  the notation $\child{i}{u}$ denotes $u$'s $i$th child;
  $\depth(u)$ denotes the depth
  (the length of the corresponding codeword);
  $\parent(u)$ denotes the parent.

  The {\em cost $c(T)$} of such a tree is the sum of the depths
  of the terminals --- also called the   
  {\em external weighted path length}  of the tree. 

  A {\em proper} tree is a tree in which every non-terminal
  has at least two children.
\end{definitions}
The goal is to find an optimal tree with $n$ terminals.
It is easy to see that some optimal tree is proper;
thus, we restrict our attention to proper trees.

Our basic tool for understanding the structure
of optimal trees is a swapping argument.
For example, in any proper optimal tree,
no non-terminal is deeper than any terminal.
Otherwise, the terminal and the subtree
rooted at the non-terminal could be swapped,
decreasing the average depth of the terminals.

We use a swapping argument to prove that 
an optimal proper tree has the following form for some $m$.
The non-terminals are the $m$ shallowest (i.e., least-depth) nodes
of the infinite tree,
while the terminals are the $n$ shallowest available
children of these nodes in the infinite tree.
We call such a tree {\em shallow};
here is the precise definition:
\begin{definition}
  A tree $T$ is {\em shallow} provided that
  \begin{itemize}
  \item[(i)] for any non-terminal $u\in T$
    and any node $w$ (not necessarily in $T$) that is not a non-terminal,
    $\depth(u) \le \depth(w)$ and
  \item[(ii)] for any terminal $u\in T$
    and any node $w$ that is not in $T$
    but is a child of a non-terminal,
    $\depth(u) \le \depth(w)$.
  \end{itemize}
\end{definition}
Note that a non-terminal of an (improper) shallow tree might have no children
in the tree.  This is why we refer to ``terminal'' and ``non-terminal'' nodes
in place of the more common ``internal nodes'' and ``leaves''.

As a simple example consider the basic binary tree; $r=2,$ $c_1=c_2=1.$
A proper binary tree $T$ will be shallow if and only if there is some depth $l$
such that (a) every node $u$ in the infinite tree with $\depth(u) < l$
is a non-terminal in $T$ and (b) all terminals of $T$
are on levels $l$ and $l+1.$ Conditions (a) and (b) are necessary and
sufficient conditions for $T$ to have minimum external path length among all
binary trees with the same number of leaves, see 
e.g.,~\cite[\S 5.3.1]{Knuth-73}. So,
a binary tree has minimum external path length for its number of leaves if and
only if it is shallow.  
For example, the binary tree on the left of Figure \ref{comparisons} has
minimum external path length among all trees with $6$ leaves because it
fulfills conditions (a) and (b) with $l = 2.$ 
As we will see later, though,
for most values of $r$ and $c_i$ shallowness 
alone does not imply optimality.
However, if a shallow tree has the right number of non-terminals,
then it is optimal:
\begin{lemma}
  \label{shallow lemma}
  Let $m^*$ be the minimum number of non-terminals in any optimal tree.
  Then any shallow tree with $m^*$ non-terminals is optimal and proper.
\end{lemma}
\begin{proof}
  Fix a shallow tree $T$ with $m^*$ non-terminals.
  We will show the existence of an optimal tree
  with the same non-terminals as $T$.
  Since $T$ is shallow, by property (ii), this will imply $T$ is optimal.
  By the choice of $m^*$, $T$ is also proper
  (otherwise there would be an optimal proper tree with fewer non-terminals).

  It remains to show the existence of an optimal tree
  with the same non-terminals as $T$.
  Let $T^*$ be an optimal (and therefore proper) tree with $m^*$ non-terminals.
  Let $N$ and $N^*$ be the sets of non-terminals of $T$ and $T^*$,
  respectively.  If $N = N^*$ we are done.
  Otherwise, let $u$ be a minimum-depth node in $N - N^*$,
  so that $u$'s parent is in $N^*$.
  Let $u^*$ be a node in $N^* - N$.
  Note that, since $T$ is shallow, $\depth(u^*) \ge \depth(u)$,
  but that, in $T^*$, $u^*$ is a non-terminal
  (with at least two terminal descendants)
  while $u$ is either a terminal or not present.

  In $T^*$, swap the subtrees rooted at $u$ and $u^*$. 
  Specifically, make $u$ a non-terminal
  and, for each descendant $v^*$ of $u^*$,
  delete it and add the corresponding descendant $v$ of $u$.
  If $v^*$ was a terminal, make $v$ a terminal,
  otherwise make $v$ a non-terminal.
  If $u$ was a terminal, make $u^*$ a terminal,
  otherwise delete $u^*$.
  Call the resulting tree $T'$.

  From $\depth(u^*) \ge \depth(u)$ it follows that $c(T') \le c(T^*)$.
  Thus, $T'$ is also optimal.
  Note that $T'$ shares one more non-terminal with $T$ than does $T^*$.
  Thus, repeated swapping produces an optimal tree
  with the same non-terminals as $T$.
\end{proof}

Note that $m^* \ge (n-1)/(r-1)$, since each node has degree at most $r$.

\begin{corollary}
  \label{sequence corollary}
  Let $\lowm = \lceil (n-1)/(r-1) \rceil$.
  Let $\langle T_{\lowm}, T_{\lowm+1}, T_{\lowm+2}, ... \rangle$
  be any sequence of shallow trees 
  such that for each $m$, $T_m$ has $m$ non-terminals.
  Then one of the $T_m$ is proper and optimal.
\end{corollary}

The algorithm generates a sequence of shallow trees as above
and returns the one which has minimum cost.
The lemma guarantees that this tree will be optimal.
The rest of the paper is devoted to examining
the properties of shallow trees which enable the enumeration of
the proper shallow trees in $O(n \log^2 r)$ time.

\begin{figure}[t]
  \centerline{\psfig{figure=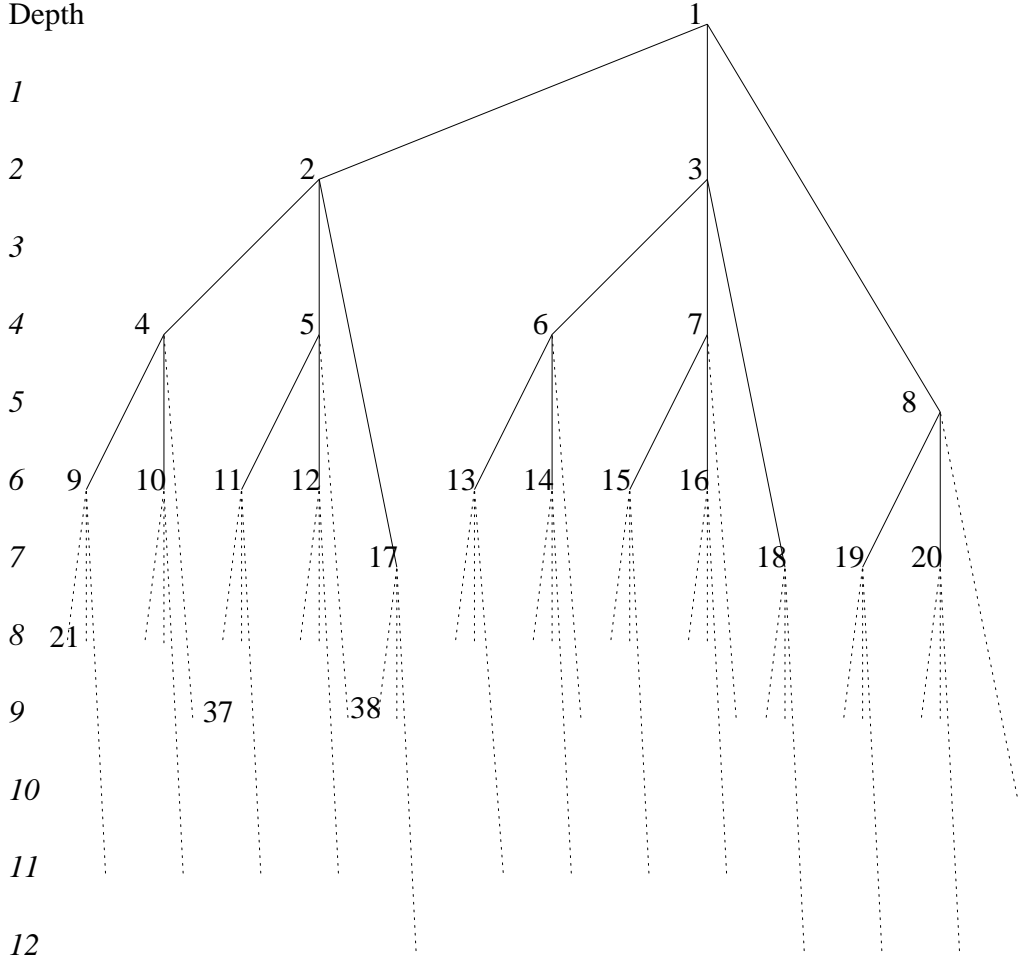,height=5in}}
  \caption{The top of a labeled infinite tree with $r=3,$
    $c_1 =2,$ $c_2=2,$ and $c_3=5.$}
  \label{infinite tree}
  \hrule
  \smallskip
\end{figure}

\subsection{Defining the Trees}  \label{tree def sec}

\paragraph{Ordering the nodes. }
Label the nodes of the infinite tree as $1,2,3,\ldots,$
in order of increasing depth.
Break ties arbitrarily,
except that 
if two nodes $u$ and $w$ are of equal depth,
and both are $i$th children of their respective parents,
and $\parent(u) < \parent(w)$,
then let $u < w $ (this is needed for Lemma \ref{child lemma}).
For the sake of notation, 
    identify each node with its label,
so that $1$ is the root, $2$ is a minimum-depth child of the root, etc.
Figure \ref{infinite tree} illustrates the top section of such a labeling
for $r=3,$ $c_1=2,$ $c_2=2,$ and $c_3=5.$  
These values of $r$ and $c_j$ are the ones we use in all later examples.

\begin{figure}
  $$
  \begin{array}{ccc}
    T_{5} &\quad& T_{6}\\
    \psfig{figure=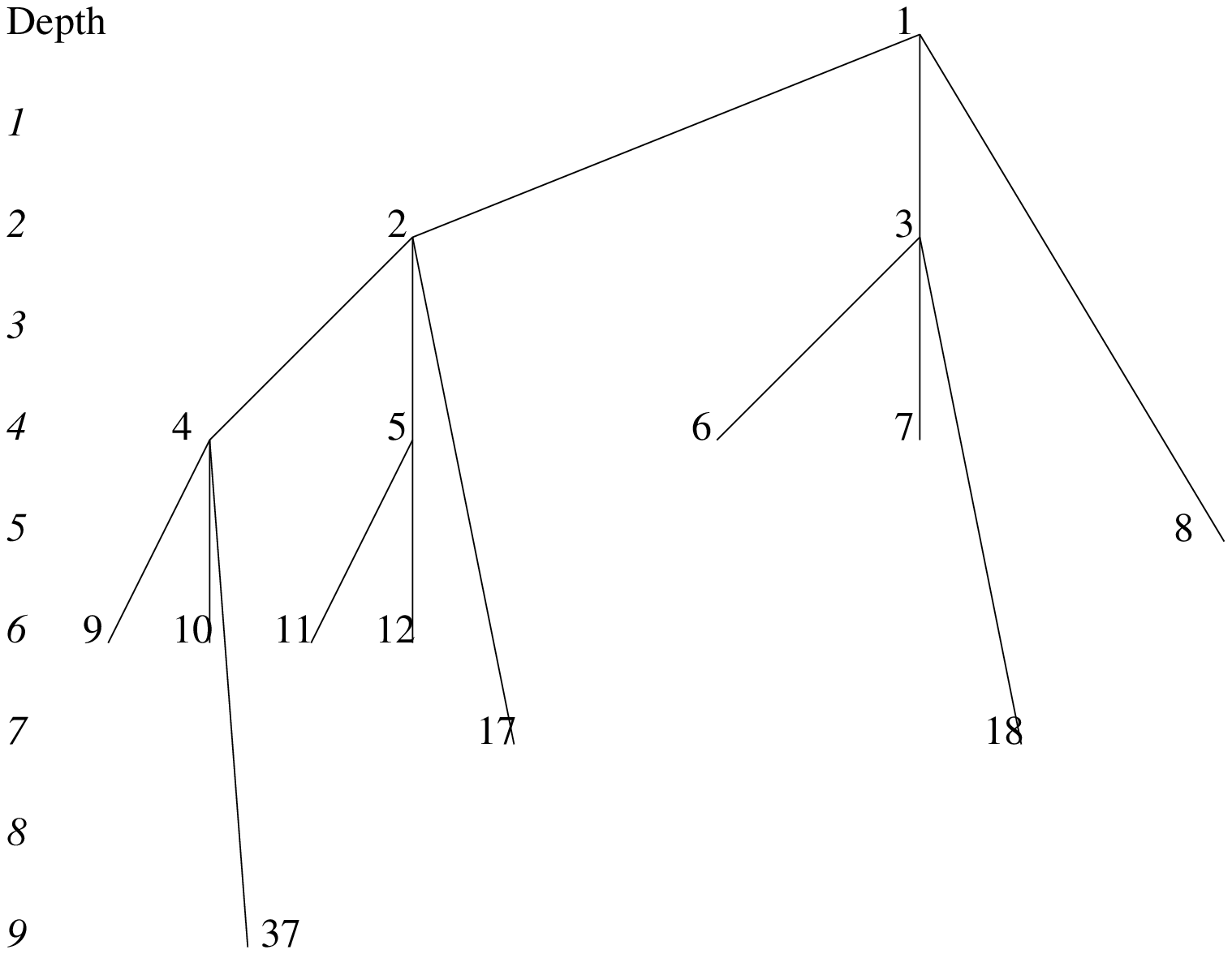,width=2.15in} & & \psfig{figure=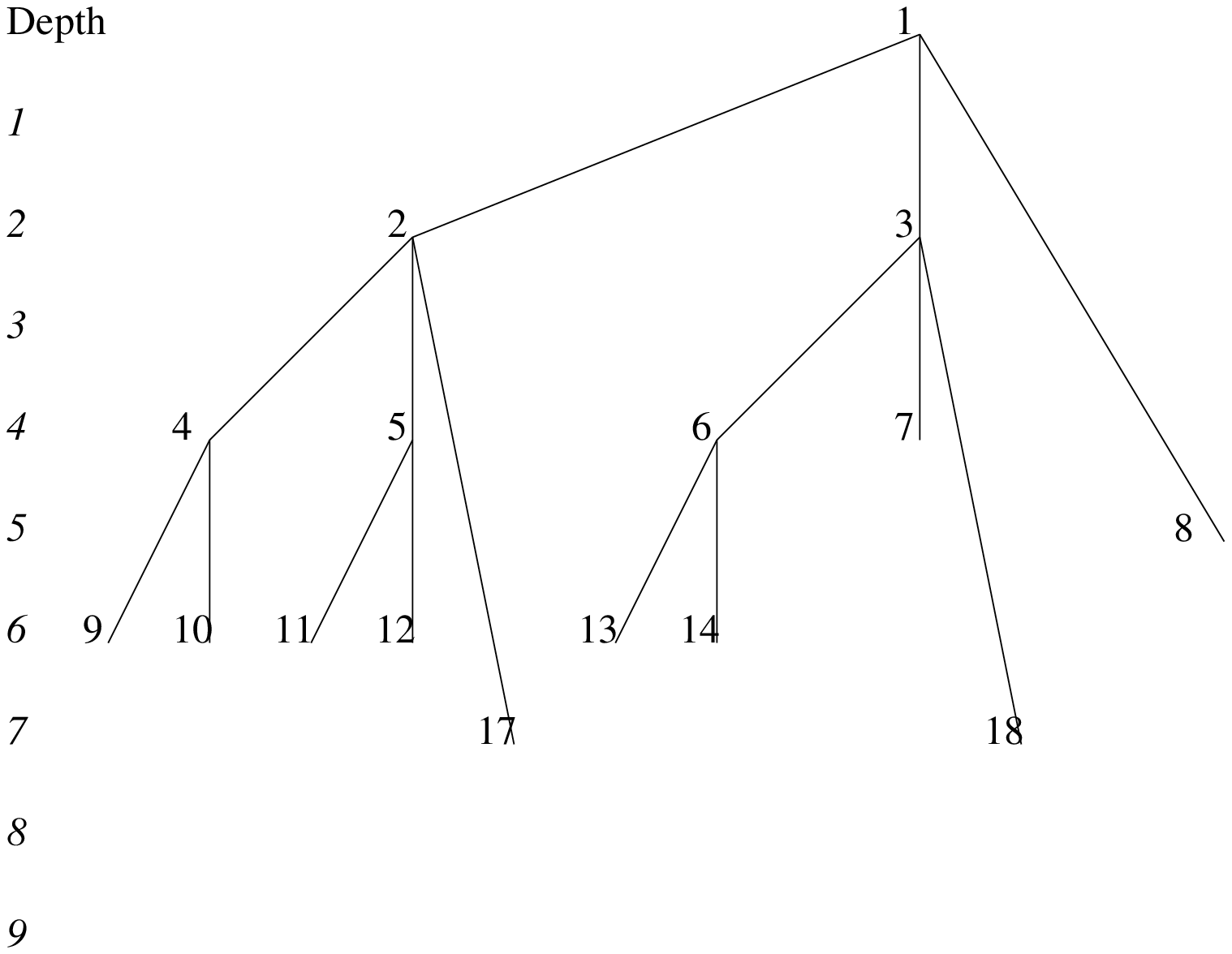,width=2.15in}\\
    \begin{array}{c|ccc}            
      i              & 1  &  2 & 3\\ \hline
      \lowkey{i}     &   3&  3 & 1\\
      \highkey{i}    &   5&  5 & 4
    \end{array}
    &&
    \begin{array}{c|ccc}            
      i              & 1  &  2 & 3\\ \hline
      \lowkey{i}     &   4&  3 & 1\\
      \highkey{i}    &   6&  6& 3
    \end{array}
  \end{array}$$

  $$
  \begin{array}{ccc}
    T_{7} &\quad& T_{8}\\
    \psfig{figure=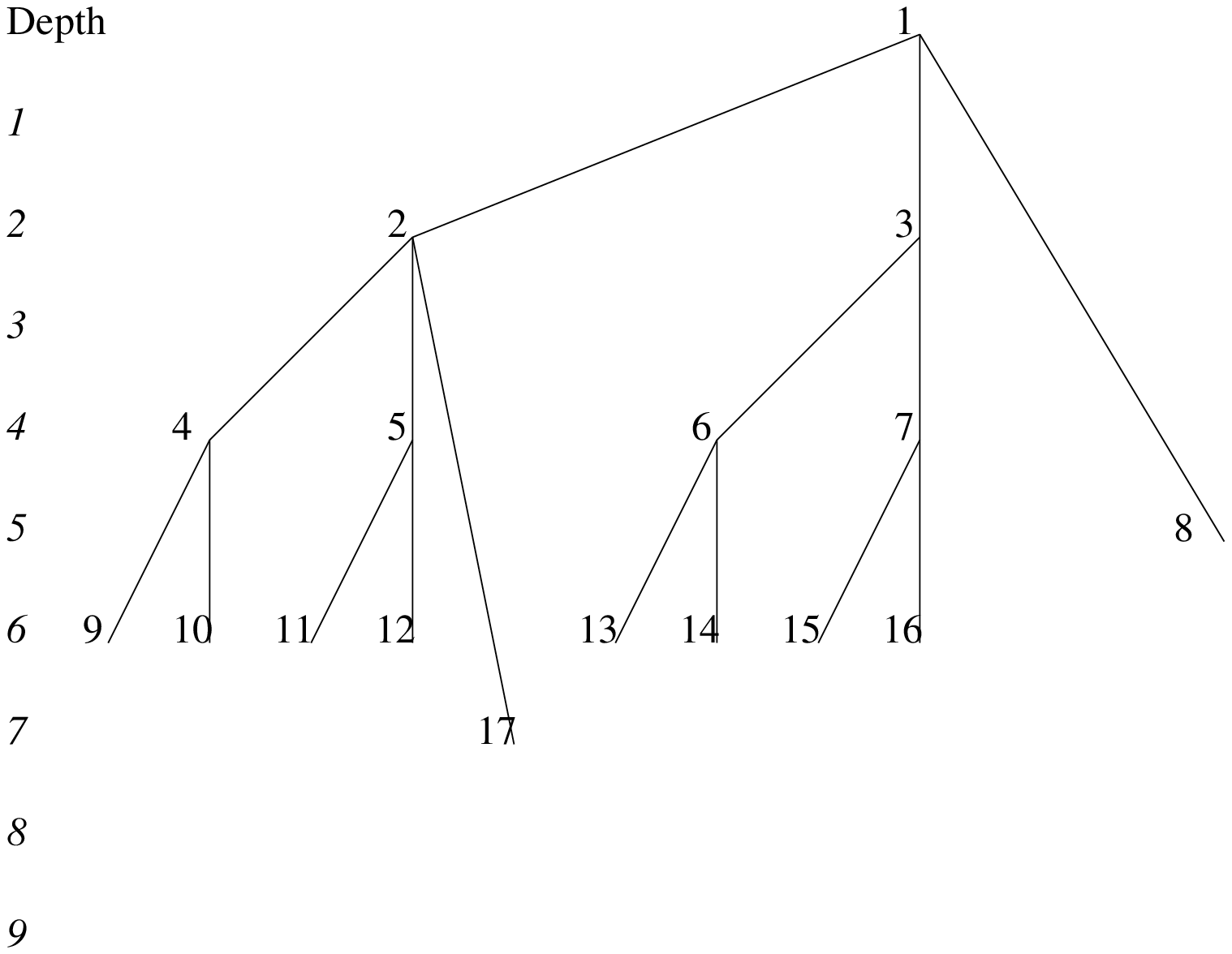,width=2.15in} && \psfig{figure=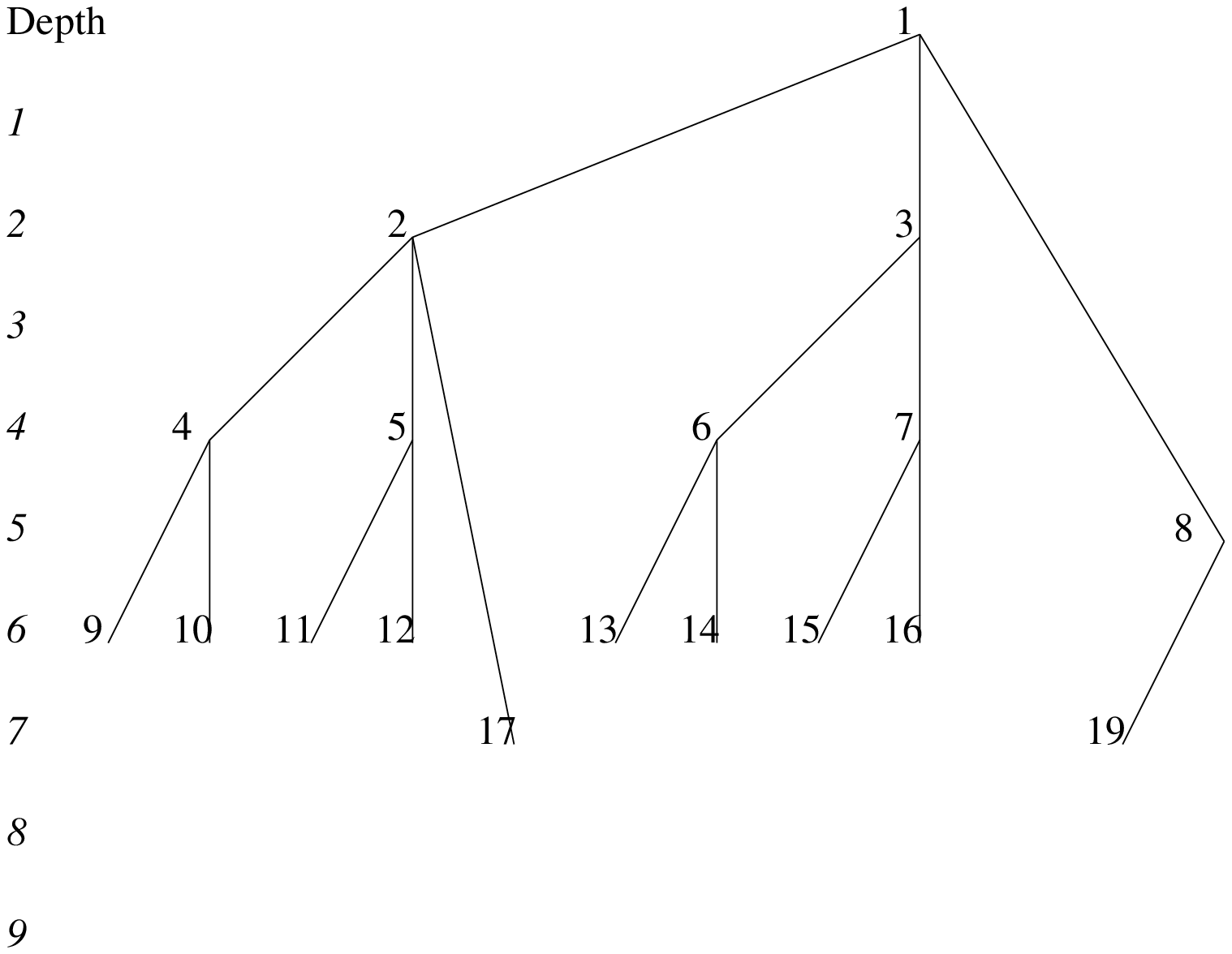,width=2.15in}\\
    \begin{array}{c|ccc}            
      i              & 1  &  2 & 3\\ \hline
      \lowkey{i}     &   4&  4 & 1\\
      \highkey{i}    &   7&  7& 2
    \end{array}
    &&
    \begin{array}{c|ccc}            
      i              & 1  &  2 & 3\\ \hline
      \lowkey{i}     &   4&  4 & 2\\
      \highkey{i}    &   8&  7& 2
    \end{array}
  \end{array}$$

  \caption{The trees $T_{5},$ $T_{6},$ $T_{7},$ and $T_{8}$ for
    $r=3,$ $c_1=2,$ $c_2=2,$  $c_3=5$ and $n=10.$ The node numbering is that of
    the previous figure. Calculating the external path lengths we find
    that $c(T_5) = 60,$ $c(T_6)=59,$ $c(T_7)=60,$ and $c(T_8) =62.$}
  \label{Tmex}
\end{figure}

\begin{definition}
  For each $m \ge \lowm$ define $T_m$ to be the tree whose non-terminals are
  $\{1,...,m\}$ and whose terminals are the minimum $n$ nodes
  among the children of $\{1,...,m\}$ in $\{m+1,m+2,...\}$.
\end{definition}
Thus, $T_m$ is the ``shallowest'' tree
with $m$ non-terminals with respect to the ordering of the nodes.
Since the ordering of the nodes respects depth, each $T_m$ is shallow.
Figure \ref{Tmex} presents $T_5,$ $T_6,$ $T_7,$ and $T_8$ for $n=10$ 
using the labeling of Figure \ref{infinite tree}.

\subsection{Relation of Successive Trees} 
Next we turn our attention to the relation of $T_{m+1}$ to $T_m$.
\begin{lemma}
  \label{sprout lemma}
  For $m \ge \lowm$, the new non-terminal (node $m+1$) in $T_{m+1}$
  is the minimum terminal of $T_m$.
\end{lemma}
\begin{proof}
  The parent of $m+1$ is in $\{1,...,m\}$,
  so $m+1$ is the minimum child of $\{1,...,m\}$ in $\{m+1,m+2,...\}$.
  The result follows from the definition of $T_m$.
\end{proof}

\begin{lemma}
  \label{level lemma}
  For $m \ge \lowm$, 
  provided the new non-terminal (node $m+1$) in $T_{m+1}$
  has at least one child,
  each terminal of $T_{m+1}$ is either a child of $m+1$ 
  or a terminal of $T_m$.
\end{lemma}
\begin{proof}
  Let node $m+1$ have $d$ children in $T_{m+1}$.
  Let $\cal C$ denote the set of children of nodes $\{1,...,m\}$
  in $\{m+1,m+2,...\}$.
  The terminals of tree $T_{m+1}$
  consist of the minimum $d$ children of node $m+1$
  together with the minimum $n-d$ nodes in ${\cal C} - \{m+1\}$.
  These $n-d$ nodes, together with node $m+1$
  (the minimum node in $\cal C$),
  are the $n-d+1$ minimum nodes in $\cal C$.
  If $d \ge 1$, then by the definition of $T_m$,
  each such node is a terminal in $T_m$.
\end{proof}

The main significance of Lemmas \ref{sprout lemma} and \ref{level lemma}
is that they will allow an efficient construction of $T_{m+1}$.
Moreover, they imply that, if $T_m$ is not proper,
neither is any subsequent tree.

\begin{lemma}
  One of the trees $\langle T_{\lowm}, T_{\lowm+1}, ..., T_\highm \rangle$
  is optimal and proper,
  where $\highm = \min \{m : T_{m+1} \mbox{ is improper}\}$.
  \label{algorithm proof lemma}
\end{lemma}
\begin{proof}
  By Lemma \ref{level lemma},
  if $T_m$ is improper, then so is $T_{m+1}$ ---
  either node $m+1$ has no children in $T_{m+1}$
  or the non-terminal in $T_m$ that had less than two children
  also has less than two children in $T_{m+1}$.
  Hence, for each $m > \highm$, tree $T_m$ is improper.
  Thus Corollary \ref{sequence corollary} implies that one of the trees
  $\langle T_{\lowm}, T_{\lowm+1}, ..., T_\highm \rangle$ 
  is proper and optimal.
\end{proof}

For $n=10,$
$\lowm = \lceil {{10 -1} \over {3-1}} \rceil = 5$ and (as shown
in Figure \ref{Tmex}) $T_8$ is improper.
The lemma then implies that one of $T_5,$ $T_6,$ or $T_7$ 
must have minimum external path length.  Calculation shows
that $T_6$ with $c(T_6) = 59$ is the optimal one.

\section{Computing the Trees}
\begin{figure}
  $$
  \begin{array}{ccc}
    T_{5} &  \qquad &  \Sprout(T_5)\\
    \psfig{figure=young3.ps,width=2.15in}
    & \qquad &
    \psfig{figure=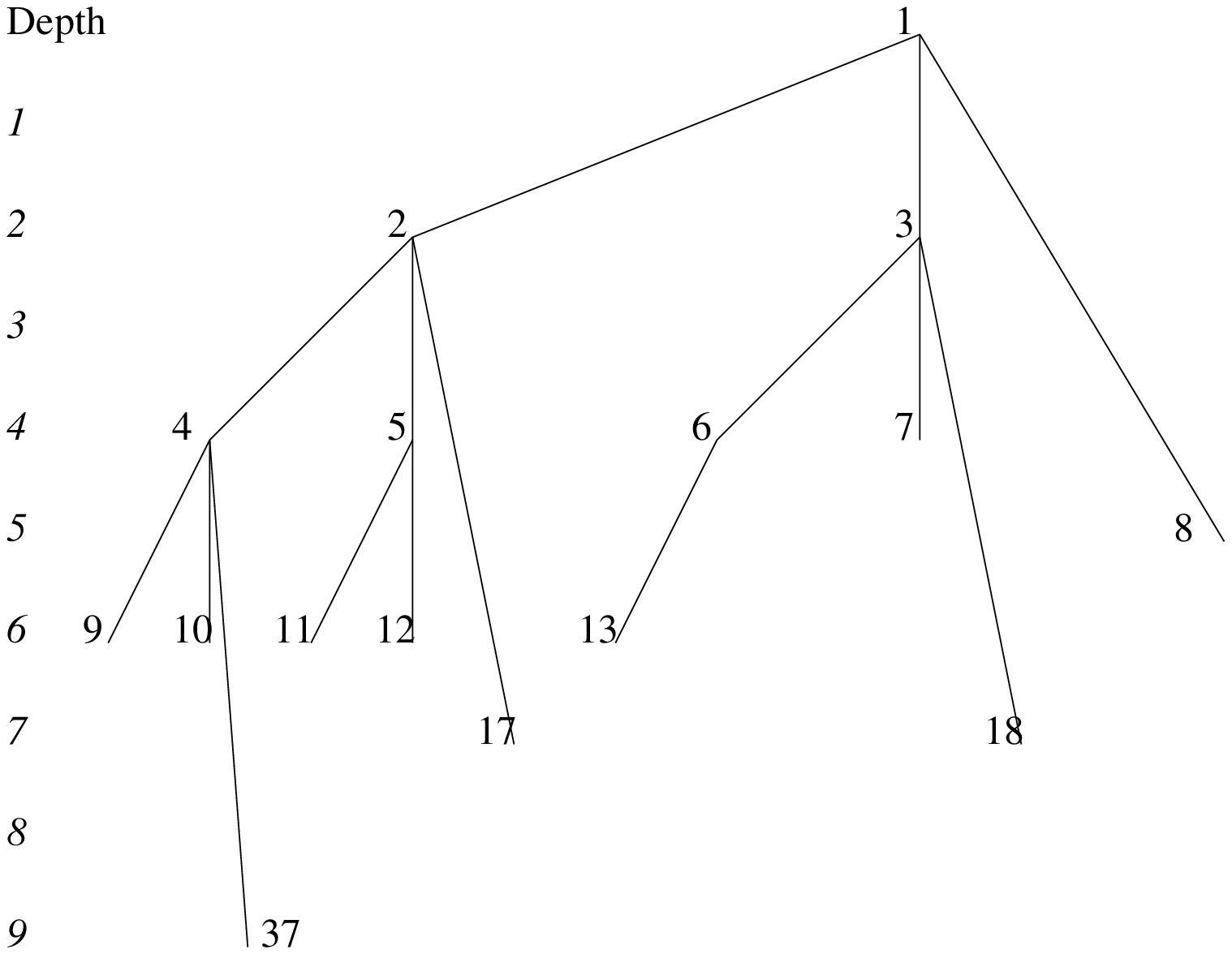,width=2.15in}\\
  \end{array}
  $$
  \bigskip
  $$
  \begin{array}{c}
    T_6 = \Level(\Sprout(T_5))\\
    \psfig{figure=young4.ps,width=2.15in}
  \end{array}
  $$
  \caption{\Sprout{}ing and \Level{}ing $T_5$ yields $T_6.$}
  \label{level and sprout picture}
\end{figure}

The algorithm uses the following two operations to compute the trees.
\begin{description}
\item To \Sprout\ a tree is to make its minimum terminal a non-terminal 
  and to add the minimum child of this non-terminal as a terminal.
\item To \Level\ a tree is to add $c$ children of the maximum non-terminal
  to the tree as terminals and to remove the $c$ largest terminals in the tree.
  The $c$ children are the minimum $c$ children not yet in the tree,
  where $c$ is maximum such that all children added 
  are less than all terminals deleted.
\end{description}
The algorithm computes the initial tree $T_{\lowm}$
then repeatedly \Sprout{}s and \Level{}s to obtain successive trees
until the tree so obtained is not proper.
Lemmas \ref{sprout lemma} and \ref{level lemma} imply that,
as long as node $m+1$ has at least one child in $T_{m+1}$ 
(it will if $T_{m+1}$ is proper),
\Sprout{}ing and \Level{}ing $T_m$ yields $T_{m+1}.$
Figure \ref{level and sprout picture} illustrates this operation.

\begin{observation}
  \label{observation}Let $m=\highm.$ 
  If node $m+1$ has at least one child in $T_{m+1}$
  then \Sprout{}ing and \Level{}ing $T_m$
  yields tree $T_{m+1}.$
  If node $m+1$ has no children in $T_{m+1}$,
  then the maximum terminal in $T_m$ is less than
  the minimum child of node $m+1$ and
  \Sprout{}ing and \Level{}ing $T_m$
  yields a tree in which non-terminal $m+1$ has one child.
  Hence, the algorithm always correctly identifies $T_\highm$
  and terminates correctly, having considered all relevant trees.
\end{observation}

To \Sprout\ requires identification 
and conversion of the minimum terminal of the current tree,
whereas to \Level\ requires identification
and replacement of (no more than $r$) maximum terminals
by children of the new non-terminal.
One could identify the maximum and minimum terminals in $O(\log n)$ time
by storing all terminals in two standard priority queues
(one to detect the minimum, the other to detect the maximum).
At most $r$ terminals would be replaced in computing each tree
and, because $\highm \le n-1,$ only $O(n)$ trees would be computed.
This approach yields an $O(rn\log n)$-time algorithm.

By a more careful use of the structure of the trees,
we improve this in two ways.
First, we give an amortized analysis showing that
in total, only $O(n\log r)$, rather than $O(rn)$, terminals are replaced.
Second, we show how to reduce the number of non-terminals
in each priority queue to at most $r$.
This yields an $O(n\log^2 r)$-time algorithm.

Both improvements follow from the tie-breaking condition
on the ordering of the nodes,
which guarantees that $T_m$ must have the following structure.

\begin{lemma}
  \label{child lemma}
  In any $T_m$, if $u$ and $w$ are non-terminals with $u < w$,
  and the $i$th child of $w$ is in the tree,
  then so is the $i$th child of $u$.
  If the $i$th child of $w$ is a non-terminal,
  then so is the $i$th child of $u$.
\end{lemma}
\begin{proof}
  Straightforward from the definition of $T_m$
  and the condition on breaking ties in ordering the nodes
  (in \S \ref{tree def sec}).
\end{proof}

\begin{corollary}
  \label{child corollary}
  Node $m$ has a minimum number of children among all non-terminals in $T_m.$
  \label{degree corollary}
\end{corollary}

\subsection{Only $O(n\log r)$ Replacements Total}
The number of terminals replaced while obtaining $T_m$ from $T_{m-1}$ 
is at most the number of children of non-terminal $m$ in $T_m$.
Although this might be $r$ for many $m$,
the sum of the numbers of children is $O(n\log r)$:
\begin{lemma}
  \label{dm bound lemma}
  Let $d_m$ be the number of children of non-terminal $m$ in tree $T_m$.
  Then $\sum_m d_m$ is $O(n\log r)$.
\end{lemma}
\begin{proof}
  By Corollary \ref{child corollary},
  within $T_m,$ node $m$ has the fewest children.
  The total number of children of the $m$ non-terminals is $m+n-1$.
  Thus, $d_m$ is at most the average $(m+n-1)/m = 1 + (n-1)(1/m)$.
  \begin{eqnarray*}
    \sum_{m=\lowm}^\highm d_m 
    & \le & (\highm - \lowm + 1) 
      + (n-1) \sum_{m=\lowm}^\highm 1/m \\
    & = & O(\highm - \lowm + n\log(\highm/\lowm)).
  \end{eqnarray*}
  The result follows from 
  $\lowm = \lceil {{n-1} \over {r-1 }}\rceil$ 
  and $\highm \le n-1.$

\end{proof}

\subsection{Limiting the Relevant Terminals} 
\label{terminals section}
To reduce the number of terminals that must be considered 
in finding the minimum and maximum terminals,
    we partition the terminals into $r$ groups.
    The $i$th group consists of the terminals that are $i$th children
    $(i=1,...,r)$.
\begin{lemma}
  \label{interval lemma}
  In any $T_m$, for any $i$, 
  the set of non-terminals whose $i$th children are terminals 
  is of the form $\{u_i,u_i+1,...,w_i\}$ for some $u_i$ and $w_i$.
  The minimum among terminals that are $i$th children
  is $\child{i}{u_i}$ (the $i$th child of $u_i$).
  The maximum among these terminals is $\child{i}{w_i}$.
\end{lemma}
\begin{proof}
  A straightforward consequence of Lemma \ref{child lemma}.
\end{proof}

Figure \ref{Tmex} presents $u_i$ and $w_i$ for the trees
$T_5,$ $T_6,$ $T_7,$ and $T_8$ when $n=10.$

This lemma implies that the minimum terminal in $T_m$
is the minimum among $\{\child{i}{u_i} : i =1,\ldots,r\}$.
Our algorithm finds the minimum terminal in $T$ by maintaining
these $r$ particular children (rather than all $n$ terminals) 
in a priority queue.
This reduces the cost of finding the minimum from $O(\log n)$ to
$O(\log r).$  Similarly the algorithm finds the maximum terminal
in $O(\log r)$ time by maintaining $\{\child{i}{w_i} : i =1,\ldots,r\}$
in an additional priority queue.

\begin{observation}\footnote{
    This observation is due to R. Fleischer. 
    }
  \label{unimodality}
  As an aside, one can prove using Lemma \ref{interval lemma} that,
  for any $m$ such that $\lowm < m < \highm$,
      $c(T_{m+1})-c(T_m) \ge c(T_m)-c(T_{m-1})$.
  That is, the sequence of tree costs is unimodal.
  To prove this, consider building $T_{m+1}$ from $T_m$.
  $\Sprout$ing increases the cost by $c_1$;
  $\Level$ing decreases the cost with each swap.
  For each swap in building $T_{m+1}$ from $T_m$,
  one can show there was a corresponding swap in building $T_m$ from $T_{m-1}$
  and that the decrease in cost (from $T_{m}$ to $T_{m+1}$) due to the former
  is bounded by the decrease in cost (from $T_{m-1}$ to $T_m$)
  due to the latter.
  Thus, in practice the algorithm could be modified to stop 
  and return $T_{m-1}$ when $c(T_m) \ge c(T_{m-1})$.
\end{observation}

\subsection{The Algorithm in Detail}
\label{details section}
The full algorithm has two distinct phases.  The first phase constructs the
base tree $T_\lowm.$ The second phase starts with $T_\lowm$ and, by
\Sprout{}ing and \Level{}ing, iteratively constructs the sequence of
shallow trees 
$$\langle T_{\lowm}, T_{\lowm+1}, T_{\lowm+2}, ..., T_\highm\rangle$$ 
and returns one which has smallest external path
length.   $T_\highm$ is the last proper tree in the sequence, i.e.,
$T_{\highm+1}$ is improper.
Lemma \ref{algorithm proof lemma} guarantees that the algorithm
returns an optimal tree.  We now describe how to implement the
first part of the algorithm in
$O(n \log r)$ time and  the second in $O(n \log^2 r)$ time; the full
algorithm therefore runs in $O(n \log^2 r)$ time.

The skeleton of the final algorithm is shown in Fig.~\ref{alg fig}.
Procedure $\CreateTlowm$ creates tree $T_\lowm,$
the variable $\C$  contains the external path length of  current tree $T_m$
and $\Deg$ contains the number of children of node $m$ in  tree $T_m.$
As presented, the algorithm computes only 
the cost of an optimal tree.  
It can easily be modified to compute the actual tree.
Note that to check that the current tree $T_m$ is proper,
by Observation \ref{observation} and Corollary \ref{degree corollary},
it suffices to check that non-terminal $m$ has at least two children.

\begin{figure}[hbt]
  \begin{tabAlgorithm}{
      $\ComputeTrees(\langle\length{1},\length{2},...,\length{r}\rangle,n)$}
    \algline $\CreateTlowm$
    \algline WHILE \= ($\Deg \ge 2$) DO 
    \algnono        \> \CMT{Compute $T_{\M+1}$ from $T_{\M}$} 
    \algline        \> $\Sprout(T)$
    \algline        \> $\Level(T)$
    \algline        \> $\CMin \leftarrow \min\{\C,\CMin\}$
    \algline RETURN $\CMin$
  \end{tabAlgorithm}
  \caption{Algorithm to find an optimal variable-length prefix code}
  \label{alg fig}
\end{figure}

The routines \Sprout\ and \Level\ 
are shown in Figure \ref{sprout and level fig}.

\begin{figure}[htb]
  \begin{tabAlgorithm}{$\Sprout(T)$}
    \algnono \CMT{Make the minimum terminal a non-terminal}
    \algline \>      $\M \leftarrow \M+1;$
    \algline \> Let $\child{i}{\lowkey{i}}$ 
                be the minimum terminal in \lowheap.
    \algline \> $\D{\M} \leftarrow \D{\lowkey{i}}+\length{i}$;
                $\lowkey{i} \leftarrow \lowkey{i} + 1$;
                $\UpdateHeaps(T,i)$
    \algline \> $\C \leftarrow \C - \D{\M}$;
                $\Deg \leftarrow 0$;
    \algnono \CMT{Add smallest child as a terminal}
    \algline \> $\AddTerminal(T)$
  \end{tabAlgorithm}
  \vspace{0.1in}
  \begin{tabAlgorithm}{$\Level(T)$}
    \algline WHILE \= (\=$\Deg < r$ and $\child{\Deg+1}{\M}$ is less than
    \algnono  \> \> the max.~terminal $\child{i}{\highkey{i}}$ in \highheap) DO
    \algline  \> $\AddTerminal(T)$
    \algnono  \> \CMT{Delete the maximum terminal}
    \algline  \>  $\C \leftarrow \C-(\D{\highkey{i}}+\length{i})$
    \algline  \>  $\highkey{i} \leftarrow \highkey{i} - 1$;
                  $\UpdateHeaps(T,i)$
  \end{tabAlgorithm}
  \vspace{0.1in}
  \begin{tabAlgorithm}{$\AddTerminal(T)$}
    \algline  $\Deg \leftarrow \Deg+1$; 
              $\C \leftarrow \C + \D{\M} + \length{\Deg}$;
                 
    \algline $\highkey{\Deg} \leftarrow \M$; $\UpdateHeaps(T,\Deg)$
  \end{tabAlgorithm}
  \caption{The Operations \Sprout\ and \Level.}
  \label{sprout and level fig}
\end{figure}

Recall that the nodes of the infinite tree are labeled in order of increasing
depth with ties broken arbitrarily except for the requirement
that if $u$ and $v$ are both of equal depth and both are $i$th children of
their respective parents, then $u < v$ iff $\parent(u) < \parent(v).$
Depending upon $c_1,c_2,\ldots,c_r,$ 
there may be many such labelings.  The algorithm we present
breaks ties lexicographically --- suppose $u$ and $v$ have the
same depth and let $u=\child{i}{u'}$ and $v = \child{j}{v'}$;
then $u < v$ iff $u' < v'$ (or $u' = v'$ and $i < j$).  Figure 
\ref{infinite tree}
illustrates this labeling for $r=3,$ $c_1 =2,$ $c_2 = 2,$ and $c_3 =5.$
The sequence of shallow trees is fully determined by this labelling.
Figure \ref{Tmex} illustrates
the shallow trees with $10$ non-terminals for these $r$ and $c$ values.

The algorithm represents the current tree $T_m$ 
with the following data structures:

\begin{description}
\item[$\n$] --- The number of terminals.
\item[$\M$] --- The number of non-terminals.
  Also the rank of the maximum non-terminal.
\item[$\C$] --- The sum of the depths of the terminals.
\item[$\Deg$] --- The number of children of non-terminal $\M$.
\item[{$\D{u}$}] --- The depth of each non-terminal $u$.
\item[{$\lowkey{i}$}] --- The rank of the minimum non-terminal (if any)
  whose $i$th child is a terminal ($1 \le i \le r$).
\item[{$\highkey{i}$}] --- The rank of the maximum non-terminal (if any)
  whose $i$th child is a terminal ($1 \le i \le r$).
  If no non-terminal has a terminal $i$th child,
  then $\lowkey{i} > \highkey{i}$.
\item[\lowheap] --- A priority queue for finding the minimum terminal. \\
  Contains $\{\child{i}{\lowkey{i}} : \lowkey{i} \le \highkey{i}\}$.
\item[\highheap] --- A priority queue for finding the maximum terminal. \\
  Contains $\{\child{i}{\highkey{i}} : \lowkey{i} \le \highkey{i}\}$.
\end{description}

For an example refer back to Figure \ref{Tmex}. Tree $T_6$ has
\begin{displaymath}
  \begin{array}{c}
    N=10, \qquad C =59, \qquad \Deg=2,
    \\ \\
    D[1] =0, \qquad 
    D[2] = 2, \qquad
    D[3] = 3, \qquad
    D[4] = 4, \qquad
    D[5] = 4, \qquad
    D[6] = 4,
    \\ \\
    \lowkey{1}=4, \qquad
    \lowkey{2}=3, \qquad
    \lowkey{3}=1, \qquad
    \highkey{1}= 6, \qquad
    \highkey{2}=6, \qquad
    \highkey{3}=3
    \\ \\
    \lowheap = \{\child{1}{4}, \child{2}{3}, \child{3}{1}\},
    \\ \\
    \highheap = \{\child{1}{6}, \child{2}{6}, \child{3}{3}\}.
  \end{array}
\end{displaymath}

The priority queues are maintained as follows.
In general, a terminal in $T_m$ can have rank (label) 
arbitrarily larger than $m$.  The algorithm explicitly maintains 
the ranks and depths of the $m$ non-terminals in the current tree;
the algorithm compares the ranks of terminals in the priority queues
via the ranks and depths of their (non-terminal) parents.
When $\lowkey{i}$ or $\highkey{i}$ changes to reflect a new current tree,
the queues are updated by the following routine:

\begin{tabAlgorithm}{$\UpdateHeaps(T,i)$}
  \algline IF \= ($\lowkey{i} \le \highkey{i}$) THEN
  \algline        \> Update $\child{i}{\lowkey{i}}$ in \lowheap\ 
                  and $\child{i}{\highkey{i}}$ in \highheap\
  \algnono        \> to maintain the queues' invariants.
  \algline ELSE Delete both nodes from their respective queues.
\end{tabAlgorithm}

Line 2 replaces the old $\child{i}{\lowkey{i}}$ in $\lowheap$
($\child{i}{\highkey{i}}$ in $\highheap$) by the new one
when $\lowkey{i}$ ($\highkey{i}$) changes.
Line 3 will only be executed if 
$\child{i}{\lowkey{i}} > \child{i}{\highkey{i}}$, which will only happen
if the tree no longer contains any $i$th child as a terminal.
Note that Lemmas \ref{level lemma} and \ref{child lemma} imply that
if, for some $i$ and $T_m$,
no non-terminal has an $i$th child in $T_m$,
then no non-terminal has an $i$th child in $T_{m+1}$.

\paragraph{Construction of the First Tree.}
Tree $T_\lowm$  has a simple structure. Its non-terminals are the
nodes $\langle 1,2,\ldots, \lowm\rangle.$
Its terminals are  the  $n$ shallowest children of nodes
$\langle 1,2,\ldots, \lowm \rangle.$

To construct $T_\lowm$
we assume that $n > r,$ 
otherwise $T_\lowm$ is simply the root and its first $n$ children.
For $1 \le m < \lowm$, 
define $T_m$ to be the tree with non-terminals $\{1,...,m\}$
and {\em all} of the $(r-1)m+1$ children of $\{1,...,m\}$ as terminals.
The proof of Lemma \ref{sprout lemma}
generalizes easily to these trees;  node $m+1$ is the minimum terminal
of $T_m.$ 

\begin{figure}[htb]
  \begin{tabAlgorithm}{
      $\CreateTlowm(T)$}
      \algnono \CMT{Create $T_1$}
    \algline \> $\lowm = \lceil { { n-1} \over {r-1}}\rceil;$
        $\D{1} \leftarrow 0;$  $\C = \sum_{i=1}^{\min \{r,n\} } \length{i};$
    \algline \> CREATE $\lowheap;$ CREATE $\highheap;$
    \algline \> FOR \= $i = 1 \mbox{ to } \min\{r,n\}$ DO
    \algline \>     \> $\lowkey{i} \leftarrow \highkey{i} \leftarrow 1$;    
                    $\UpdateHeaps(T,i);$    
    \algnono

      \algnono \CMT{Create $\langle T_2,T_3,\ldots,T_{\lowm-1}\rangle$}
    \algline \> FOR \= $\M = 2 \mbox{ to } (\lowm-1)$ DO
    \algline \>     \> Let $\child{i}{\lowkey{i}}$ 
                    be the minimum terminal in \lowheap.
    \algline \>     \> $\D{\M} \leftarrow \D{\lowkey{i}}+\length{i}$;
                       $\lowkey{i} \leftarrow \lowkey{i} + 1$;                 
                       $\UpdateHeaps(T,i);$    
    \algline \>     \> FOR  \= $j = 1 \mbox{ to } r$ DO
    \algline \>     \>      \> $\highkey{j} \leftarrow \M$;
                            $\UpdateHeaps(T,j);$    
    \algline \>     \> $\C \leftarrow \C-\D{\M}
                         +\sum_{j=1}^r (\D{\M} +\length{j});$
    \algnono
    \algnono \CMT{Create $T_\lowm$}
    \algline \> $\M = \lowm;$ $\Delta = n - (r-1)(\lowm-1);$
    \algline \> Let $\child{i}{\lowkey{i}}$ 
                be the minimum terminal in \lowheap.
    \algline \> $\D{\M} \leftarrow \D{\lowkey{i}}+\length{i}$;
             $\lowkey{i} \leftarrow \lowkey{i} + 1$;            
             $\UpdateHeaps(T,i)$ 
    \algline \> FOR \= $j = 1 \mbox{ to } \Delta$ DO
    \algline \>     \>  $\highkey{j} \leftarrow \M$; $\UpdateHeaps(T,j);$
    \algline \> $\C \leftarrow \C-\D{\M}
                  +\sum_{j=1}^\Delta (\D{\M} +\length{j});$
    \algline \> $\Deg = \Delta;$
    \algline \> $\Level(T);$
        
  \end{tabAlgorithm}
  \caption{Operation $\CreateTlowm.$}
  \label{create first}
\end{figure}

The tree $T_1$ is easy to construct.  It is the tree with $1$ root and
$r$ children.  Inductively construct the tree $T_{m}$ from the tree
$T_{m-1},$ $m< \lowm$ as follows: find the minimum terminal in
$T_m$ by taking the minimum terminal out of $\lowheap.$ Label this node
$m,$  make it a non-terminal, and add all of its children to
$T_{m}$ as terminals.  The details are shown in Fig.~\ref{create first}.

Finally, construct $T_\lowm$ from $T_{\lowm -1}$ by making the lowest
terminal of $T_{\lowm -1}$ into node $\lowm.$  Add the $n-(r-1)(\lowm-1)$
minimum children of node $\lowm$ as terminals bringing the total number
of terminals in the current tree to $n.$ Level the resulting tree.

Since only $O(n/r)$ trees are constructed while computing $T_\lowm$
and each tree can be constructed from the previous tree in $O(r \log r)$ time,
the time required to compute $T_\lowm$ is $O(n\log r)$.  
(If desired, the time for each tree $T_m$ with $m < \lowm$
can be reduced to $O(\log r)$,
because maximum terminals are not replaced in constructing such a tree.)

\paragraph{Construction of the Remaining Trees.}  
The algorithm constructs the sequence of trees
$$\langle T_{\lowm}, T_{\lowm+1}, T_{\lowm+2}, ..., T_\highm\rangle$$
as described previously.  Tree $T_{m}$ is found 
by \Sprout{}ing and then \Level{}ing  its predecessor $T_{m-1}.$  
The cost is
$O(d_m\log r)$ time, 
where $d_m$ is the number of children of the new non-terminal $m$ in $T_m.$
By Lemma \ref{dm bound lemma} this part of the algorithm runs in
$O\left( \left(\sum_m d_m\right) \log r\right) = O(n \log^2 r)$
time.

\bigskip
\par\noindent{\em Acknowledgements:} 
The authors would like to thank Dr.
Jacob Ecco for introducing us to the Morse Code puzzle which sparked
this investigation.  They would also like to thank Siu Ngan Choi and Rudolf 
Fleischer
(who made Observation \ref{unimodality}
--- the unimodality of the tree costs)
for their careful reading of an earlier manuscript and 
subsequent comments.

\end{document}